\documentclass[twocolumn]{aastex631}

\usepackage{enumitem}
\setlist[itemize]{leftmargin=*}

\newcommand{\tl}{\texttt{TESS\_localize}}

\defcitealias{Dorn-Wallenstein2019}{DW19}
\defcitealias{Dorn-Wallenstein2020}{DW20}
\defcitealias{Dorn-Wallenstein2022}{DW22}

\shorttitle{Contamination in TESS light curves}
\shortauthors{Pedersen \& Bell}

\begin{document}

\title{\Large Contamination in TESS light curves: The case of the Fast Yellow Pulsating Supergiants}

\email{may.pedersen@sydney.edu.au}

\author[0000-0002-7950-0061]{May G. Pedersen}
\affiliation{Sydney Institute for Astronomy, School of Physics, University of Sydney NSW 2006, Australia}
\affiliation{Kavli Institute for Theoretical Physics, Kohn Hall, University of California, 
Santa Barbara, CA 93106, USA}

\author[0000-0002-0656-032X]{Keaton J.\ Bell}
\affiliation{Department of Physics, Queens College, City University of New York, Flushing, NY-11367, USA}

\begin{abstract}

Given its large plate scale of 21\arcsec\ pixel$^{-1}$, analyses of data from the TESS space telescope must be wary of source confusion from blended light curves, which creates the potential to attribute observed photometric variability to the wrong astrophysical source. We explore the impact of light curve contamination on the detection of fast yellow pulsating supergiant (FYPS) stars as a case study to demonstrate the importance of confirming the source of detected signals in the TESS pixel data.
While some of the FYPS signals have already been attributed to contamination from nearby eclipsing binaries, others are suggested to be intrinsic to the supergiant stars. In this work, we carry out a detailed analysis of the TESS pixel data to fit the source locations of the dominant signals reported for 17 FYPS stars with the Python package \tl. We are able to reproduce the detections of these signals for 14 of these sources, obtaining consistent source locations for four. Three of these originate from contaminants, while the signal reported for BZ Tuc is likely a spurious frequency introduced to the light curve of this 127-day Cepheid by the data processing pipeline. Other signals are not significant enough to be localized with our methods, or have long periods that are difficult to analyze given other TESS systematics. Since no localizable signals hold up as intrinsic pulsation frequencies of the supergiant targets, we argue that unambiguous detection of pulsational variability should be obtained before FYPS are considered a new class of pulsator. 

\end{abstract}



\section{Introduction} \label{sec:intro}

The Transiting Exoplanet Survey Satellite \citep[TESS][]{TESS} is obtaining extensive time series photometry over most of the sky, enabling the study of photometric variability of nearly all types of astronomical objects. Interested in the behavior of massive evolved stars, \citet[][hereafter DW19]{Dorn-Wallenstein2019} inspected TESS light curves for variability that could improve our understanding of these objects. In addition to apparent luminous blue variables (LBVs), unprecedented variability was identified in the TESS light curves of three yellow supergiants (YSGs). The most notable of these is HD~269953, which shows multiple periodicities.

That work was later extended, utilizing the full TESS Cycle 1 dataset to investigate variability of 76 YSGs \citep[][hereafter DW20]{Dorn-Wallenstein2020}. Among these, a total of five (HD~269953, HD~269110, HD~268687, HD~269840, and HD~269902) were interpreted to be rapidly pulsating on $\sim$day timescales, suggesting that these belong to a new class of pulsator, which were named FYPS for ``fast yellow pulsating supergiants.'' If this is true, then FYPS would offer new exciting opportunities for studying and modelling the interiors of evolved massive stars. 

\begin{figure}
\center
\includegraphics[width=\linewidth]{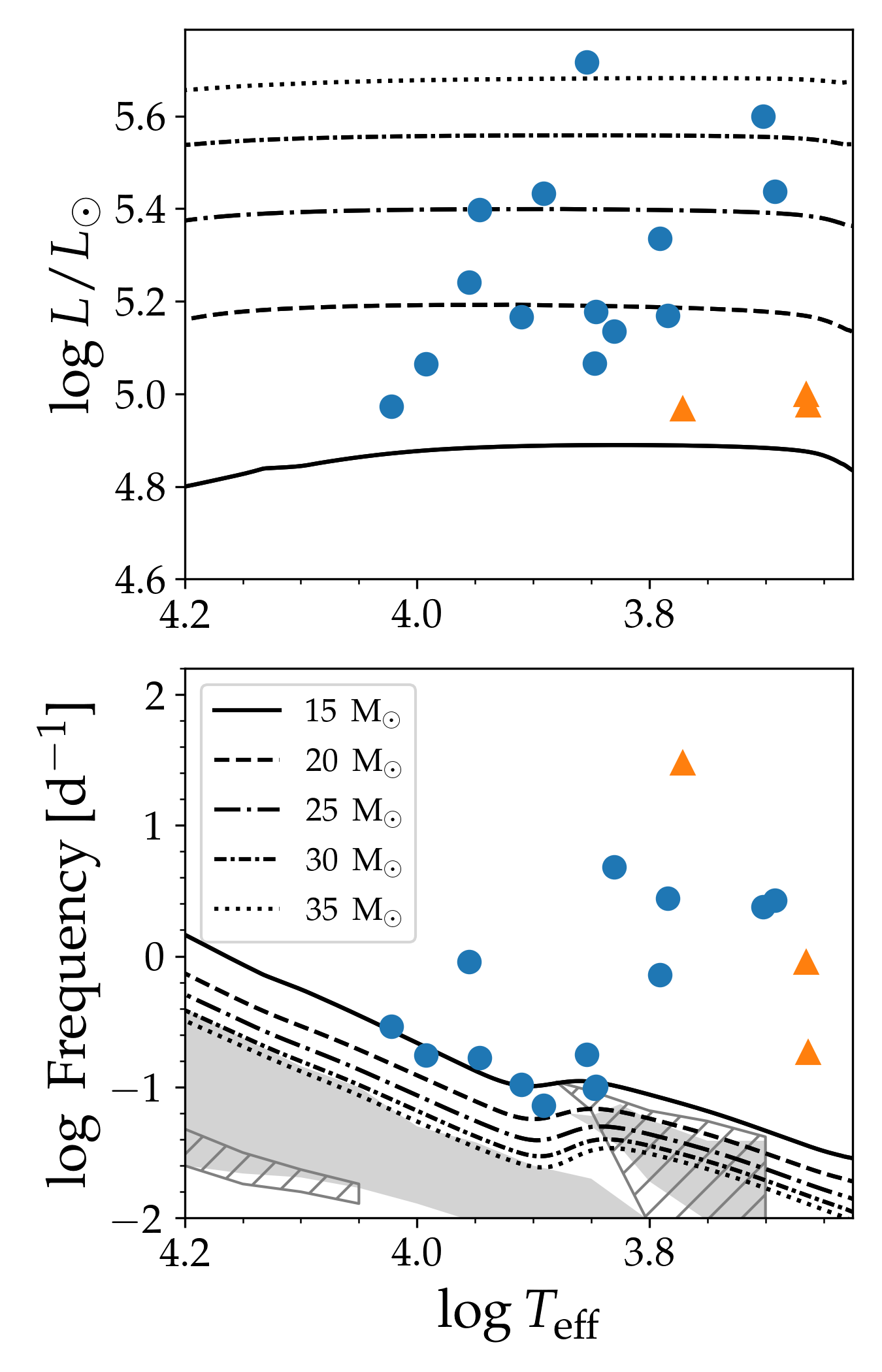}
\caption{Position of the 17 FYPS stars in the HR diagram (top) and their dominant frequency $f_0$ reported by \citetalias{Dorn-Wallenstein2022} 
as a function of effective temperature (bottom). Blue circles indicate LMC stars, whereas orange triangles correspond to SMC stars. Example evolutionary tracks computed with \texttt{MESA} for LMC composition (see Appendix~\ref{Sec:mesa}) are shown by black lines, with the corresponding evolution of the acoustic cutoff frequency shown in the bottom panel. Different line styles correspond to different initial stellar masses as indicated by the legend. The combined predicted excited frequency ranges for a 20 and 25\,M$_\odot$ star from Fig.~5 of \citet{Saio2013} at solar metallicity are shown as a function of $\log T_{\rm eff}$ for pre-RSG (hatched grey) and post-RSG (light grey) evolution, respectively.}\label{fig:acou_cutoff}
\end{figure}

Motivated by these prospects, \citet[][hereafter DW22]{Dorn-Wallenstein2022} set out to identify additional FYPS stars in the large and small Magellanic Clouds (LMC and SMC). The positions of their 17 candidate FYPS stars in the HR diagram are shown in the top panel of Fig.~\ref{fig:acou_cutoff} and compared to non-rotating stellar evolution tracks of 15-35\,M$_\odot$ stars of LMC composition. The highest-amplitude signals are shown as a function of effective temperature in the bottom panel of Fig.~\ref{fig:acou_cutoff}, and they generally have higher frequencies than the acoustic cutoff frequencies of the models shown by the black curves. Above this cutoff, pressure modes are no longer reflected and trapped inside the stellar interior, but become travelling waves and dissipate. Details of these calculations are provided in Appendix~\ref{Sec:mesa}. For comparison, we also show the combined predicted frequency ranges of excited modes of a 20 and 25\,M$_\odot$ star derived by \citet{Saio2013} before (hatched grey) and after (light grey) the red supergiant (RSG) evolutionary stage assuming solar metallicity, which the highest-amplitude FYPS frequencies also exceed.

The work by \citetalias{Dorn-Wallenstein2022} included a renewed appreciation for possible source confusion due to blending of multiple sources in the 21\arcsec\,pix$^{-1}$ TESS images. Comparing their detected frequencies to signals detected in OGLE light curves \citep{OGLE} of nearby stars, they reject that several individual signals are intrinsic to the YSGs. Two of the original five FYPS stars (HD~269110 and HD~269902) were found to be contaminated as a result, along with roughly half of their initial sample of TESS variable light curves. Remaining signals that cannot be attributed to other OGLE sources are interpreted as being intrinsic to the YSGs and used to classify these objects as FYPS. The authors argue that a statistically significant excess of variable light curves extracted for stars with $\log{L/L_\odot} > 5$ supports that FYPS are not purely the result of contamination.

We demonstrate in this work that the TESS data themselves show additional signals for the 17 FYPS stars identified by \citetalias{Dorn-Wallenstein2022} 
to originate from other contaminating sources. We utilize the Python package \tl\ \citep{Higgins2022} to fit the source locations for the strongest signals reported for these objects. Most of these signals are too weak in individual TESS sectors to be localized, but every signal that can be localized is shown to be significantly offset from the YSGs (except for BZ Tuc, which is a long-period Cepheid variable). This includes two more of the original five stars used to define the FYPS class, HD~269953 and HD~268687.
We argue that FYPS should not be considered a unique region of the pulsational H-R diagram until pulsational variability can be uniquely attributed to YSGs. 
This case study demonstrates the importance of using the TESS pixel data to assess potential contamination when interpreting any signals detected by TESS given its large plate scale.


\section{The case of HD~269953 (TIC~404850274)} \label{sec:case_example}

Expanding the sample to 17 FYPS, \citetalias{Dorn-Wallenstein2022} 
no longer provide the full prewhitening solutions as were in their previous papers. Instead they indicate the approximate frequency of the highest amplitude signal that they associate with each YSG, which they call $f_0$. This is after they reject signals that they are able to attribute to different OGLE sources. 
We detail our analysis of the prototypical FYPS star HD~269953 (TIC~404850274) here to demonstrate how we interpret their work. 

HD~269953 was one of the first YSGs found to show pulsational FYPS variability, with \citetalias{Dorn-Wallenstein2019}
reporting 14 frequencies of variation. Four frequencies are reported in \citetalias{Dorn-Wallenstein2020}: 1.593, 1.174, 1.335, and 2.671\,d$^{-1}$, in order of decreasing amplitude. Changes in the prewhitening procedure and how the additional stochastic low-frequency variability is handled is likely the reason for the change in extracted frequencies. The last two reported frequencies are a harmonic pair that are likely associated with the same physical process. Table B1 of \citetalias{Dorn-Wallenstein2022} 
reports that HD~269953 has six signals, with the 2.671\,d$^{-1}$ peak having the highest amplitude ($f_0$). The periodogram shows additional significant peaks that do not appear to have been considered by these works, some of which are of higher signal-to-noise (S/N) ratio than the extracted frequencies. Here we investigate the four signals from \citetalias{Dorn-Wallenstein2020}, which includes $f_0$ and its subharmonic. Figure~B1 in \citetalias{Dorn-Wallenstein2022}\footnote{The arXiv version {\tt 2206.11917v3} shows the frequencies associated with nearby contaminants as red, vertical, dashed lines for HD~269953 in their Fig.~B1.} implies that they find that the previously-dominant 1.593 and 1.174\,d$^{-1}$ signals can be ascribed to other OGLE sources. 

\citet{Higgins2022} developed a method that utilizes the high frequency resolution of TESS to mitigate source confusion challenges caused by its low spatial resolution. They demonstrate that the amplitudes of observed variations (modeled as sinusoids) measured by each individual pixel is proportional to the flux from the variable source recorded by that pixel. The distribution of best-fit amplitudes can be fit with the TESS pixel response function (PRF) model \citep{2010ApJ...713L..97B} to constrain the position of the variable star on the sky.  

The localization method is implemented in an open-source Python package, \tl,\footnote{\url{https://github.com/Higgins00/TESS-Localize}} that we use to analyze the signals observed in the light curve of HD~269953. The code takes a TESS Target Pixel File and observed frequencies of variability as inputs, and returns a best-fit source location. For best results, \tl\ users are encouraged to validate the quality of the output; we reject all results with fitted PRF amplitudes less than five times their reported uncertainties (insignificant detections) or a reduced $\chi^2 > 20$ for the fit (poor fit agreement; see discussion in \citealt{Higgins2022}). If the localization with the aperture used by the TESS Science Processing Operations Center (SPOC) pipeline does not return a satisfactory result, we try again with the ``autoaperture'' convenience function, which can produce better results if most of the signal is located outside the pipeline aperture. We let \tl\ attempt to automatically determine an appropriate number of principal components to use to detrend the light curves.

Besides the pair of harmonic signals, there is no reason to assume that all detected signals originate from a single source in the blended TESS photometry. We run the pair of harmonics through \tl\ together and the two other signal frequencies individually. We attempt to localize these signals in each of the 23 TESS sectors of two-minute-cadence data available between Sectors 1 and 39. Considering the fits that pass our quality criteria (the majority for each signal), each frequency of variation gets localized to a consistent but different location on the sky. All are well separated from the YSG by more than a pixel, with a fit precision of $\approx 0.1$\,pix. 

\begin{figure*}
\center
\includegraphics[width=0.90\linewidth]{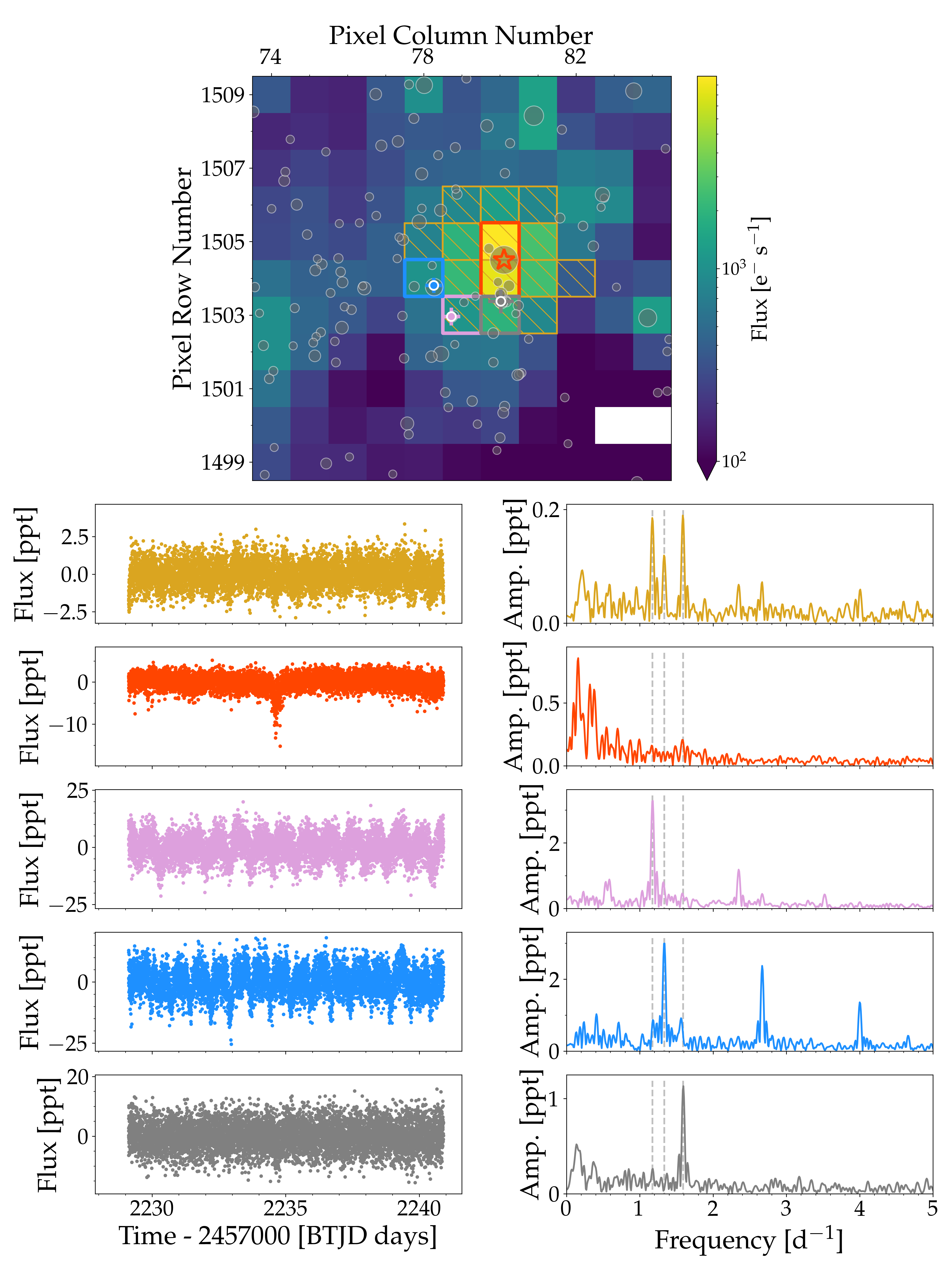}
\caption{Median TPF flux of TESS sector 34 for the YSG star HD~269953 (TIC~404850274, red star) and positions of three nearby contaminating stars shown in grey, pink, and blue. These points mark the average locations obtained for each signal across 23 sectors, with error bars representing 3\,$\sigma_{\rm std}$. Bottom panels provide the first half of each extracted light curve (left; corresponding to one spacecraft orbit) and associated periodograms (right) for the five different pixel masks indicated in the top panel. The frequency values of the three independent signals identified by \citetalias{Dorn-Wallenstein2020} 
based on the SPOC PDCSAP light curves (yellow) are shown by vertical dashed lines in the periodograms. See text for further details.}\label{fig:Cont_example1}
\end{figure*}

Figure~\ref{fig:Cont_example1} displays all the reliable source locations obtained by \tl\ plotted over the TESS Sector~34 Target Pixel File (TPF) in the top panel.
The position of HD~269953 on the TPF is indicated by the red star, while the aperture mask used to extract the 2-min cadence light curve by the SPOC pipeline is shown by the hatched yellow region.
Average fit locations corresponding to each of the three independent frequencies are presented with white circles with three different fill colors, including their 3\,$\sigma_{\rm std}$ as error bars. 
The average (RA, Dec) fit coordinate for the 1.593\,d$^{-1}$ signal shown in grey is (85.0370, -69.6726) in decimal degrees (23.7\arcsec\ from the YSG). The 1.174\,d$^{-1}$ signal (pink) is centered on (85.0160, -69.6693), and the 1.335\,d$^{-1}$ and its harmonic $f_0$ (blue) are localized to (85.0197, -69.6638) degrees (43.2\arcsec\ and 41.5\arcsec\ from the YSG, respectively). Figure~\ref{fig:Cont_example1} also shows the locations of known Gaia sources down to a $G$-magnitude of 17. 

The bottom panels of Figure~\ref{fig:Cont_example1} display the light curves and their periodograms extracted with different apertures, color-coded to match the outlined apertures in the example image. Details on how the light curves were extracted and normalized are given in Appendix~\ref{Sec:lc_extract}. The first row shows the Pre-search Data Conditioning Simple Aperture Photometry (PDCSAP) light curve and periodogram from the TESS SPOC pipeline aperture; the three main signals are marked in the periodogram with dashed vertical lines. Restricting the aperture to the two pixels containing most of the light from the target HD~269953, the next row shows the same signals to be absent. Extracting light curves from the individual pixels containing the average \tl\ fit locations for the three independent signals for the next three panels confirms where these signals are observed to be strongest. Different signals varying in amplitude differently across pixels is a telltale sign of blended variable light curves \citep{2017MNRAS.469.3802C}. 

\begin{figure*}
\center
\includegraphics[width=0.90\linewidth]{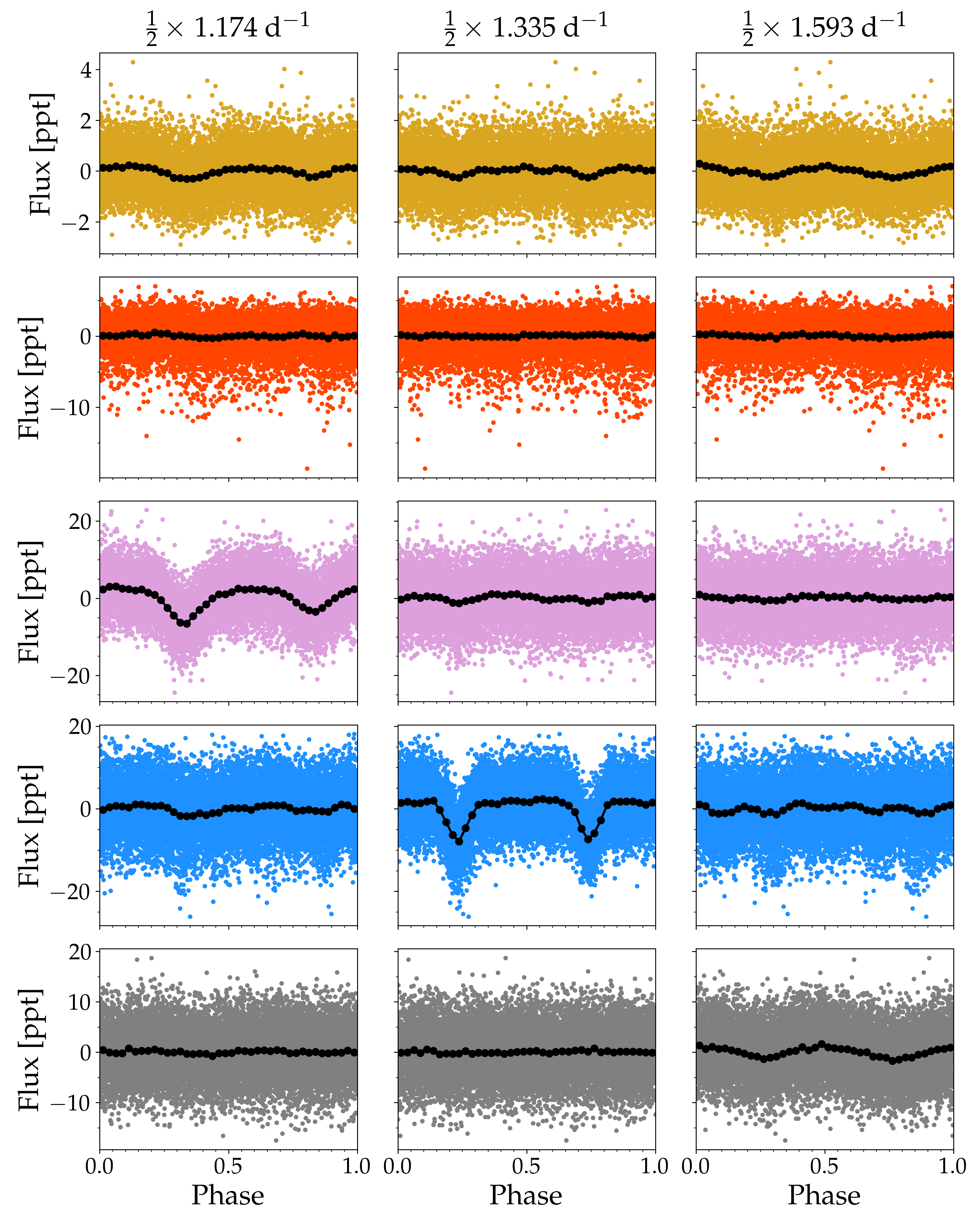}
\caption{Light curves from the left-hand panels of Fig.~\ref{fig:Cont_example1} phase folded by one half of the three independent frequencies 1.174\,d$^{-1}$ (left), 1.335\,d$^{-1}$ (center), and 1.593\,d$^{-1}$(right) identified by \citetalias{Dorn-Wallenstein2020} for HD~269953, and corresponding to the three vertical dashed lines in the right hand panels of Fig.~\ref{fig:Cont_example1}. The black dotted curves show the average flux calculated in 0.05 phase bins. We note again that \citetalias{Dorn-Wallenstein2022} had already found 1.174\,d$^{-1}$ and 1.593\,d$^{-1}$ to originate from nearby contaminants.}\label{fig:phase}
\end{figure*}

Figure~\ref{fig:phase} shows the light curves from the left hand panels of Fig.~\ref{fig:Cont_example1} phase folded by the half value of the three independent frequencies 1.174\,d$^{-1}$ (left), 1.335\,d$^{-1}$ (center), and 1.593\,d$^{-1}$ (right). As previously mentioned, the 1.174\,d$^{-1}$ and 1.593\,d$^{-1}$ signals were already associated by \citetalias{Dorn-Wallenstein2022} to nearby OGLE sources known to be eclipsing binaries (OGLE LMC-ECL-21974 and OGLE LMC-ECL-22011 with $V=15.859\,$mag and $V=15.1\,$mag, respectively), and both correspond to the second harmonic of their orbital frequencies \citep{Pawlak2016AcA....66..421P}. The eclipses can clearly be seen in the light curves centered on the two OGLE sources (pink and grey) when they are folded on their corresponding orbital frequencies, and are also visible (but weaker) in the light curves extracted from the SPOC pipeline aperture. This is also represented in the differences between the amplitudes of the signals shown in the periodograms in Fig.~\ref{fig:Cont_example1}. In comparison, the signal does not show up in the red phase-folded light curves that contains a higher fraction of light from the YSG target. 

The source location of the 1.335\,d$^{-1}$ signal coincides with the O-type star BI~265 ($V=12.375\,$mag). Phase folding the light curve by $\frac{1}{2} \times 1.335$\,d$^{-1}$ reveals two clear eclipses that are not obvious in the light curve of the YSG ($V=9.95\,$mag). While BI~265 has not previously been identified as a binary star, this result is not surprising given the expected high binary fraction for O-type stars \citep{Sana2013A&A...550A.107S,Kobulnicky2014ApJS..213...34K}. We note that while all three identified sources are significantly fainter than the YSG target ($2.4-5.9$\,mag in the V-band), their high-amplitude variability signals are still able to contaminate the light curve of the YSG.

It is clear that none of the signals reported for HD~269953 from \citetalias{Dorn-Wallenstein2020} originate from the YSG target.


\section{Localizing the FYPS signals}\label{sec:methods}

As demonstrated in detail for the FYPS prototype HD~269953 above, we can use \tl\ to fit the sky positions of signals reported to belong to FYPS pulsators. 
Most of the FYPS signals are only detectable with multiple sectors of data; since \tl\ can only fit a source location in a single sector at a time, it cannot reliably localize such weak signals.  For signals that are strong enough to be detected in individual sectors, multiple sectors of data enable multiple independent localizations of the same signals. Finding consistent source locations across multiple sectors verifies the reliability of the results, especially since images from different sectors are oriented differently on the sky. 
The $f_0$ value provided in Tables~B1 and B2 of \citetalias{Dorn-Wallenstein2022}
is the frequency of the highest-amplitude peak that they associate with each star in their sample. We attempt to recover these $f_0$ signals to full precision following a similar prewhitening methodology.

For each YSG star, we downloaded the 2-min cadence SPOC light curves from the Mikulski Archive for Space Telescopes (MAST) for all available sectors using the \texttt{lightkurve} Python package. The specific observations analyzed can be accessed via the following DOI:\dataset[10.17909/tckv-fb67]{https://doi.org/10.17909/tckv-fb67}. Following \citetalias{Dorn-Wallenstein2022}, we used the PDCSAP flux light curves, which have had corrections applied to mitigate systematic trends, and we divided the flux in each sector by its corresponding median before combining data from all available sectors. These light curves should be identical to the ones used by \citetalias{Dorn-Wallenstein2022}. Subsequently, we changed the flux units to parts-per-thousand (ppt), centered the light curves around zero, and carried out an iterative prewhitening following a similar procedure adopted by \cite{Pedersen2021NatAs...5..715P}, where the highest S/N signals are removed first using publicly available time series analysis Python packages from the IvS repository.\footnote{\url{https://github.com/IvS-KULeuven/IvSPythonRepository}} Except for  HV~829, we limited the prewhitening to frequencies below 10\,d$^{-1}$ and stopped once $S/N < 4$, calculated within a 1\,d$^{-1}$ window of the extracted frequency. This is a lower S/N threshold than is advised for general signal detection in such extensive light curves \citep{2021AcA....71..113B}, but it is useful here where we are aiming to recover signals at specific frequencies. The resulting extracted frequency lists were filtered for close frequencies using the \cite{LoumosDeeming1978} resolution criterion of $2.5/T$, where $T$ is the length of the light curve. We crossmatched this final list of frequencies with the tabulated $f_0$ values from Table~B1 and B2 of \citetalias{Dorn-Wallenstein2022}, in order to recover the reported $f_0$ values to a higher precision. 

{\movetabledown=0.5in
\tabcolsep=5pt
\begin{deluxetable*}{lccccccc}
\tablecaption{List of frequencies used to carry out the source identification with \tl\ for the 17 FYPS stars.
 \label{Tab:f0}}
\tablewidth{700pt}
\tabletypesize{\small}
\tablehead{
\colhead{TIC ID} & 
\colhead{Common name} & 
\colhead{$N_{f}^{\text{DW}}$} & 
 \colhead{$f_{0}^{\text{DW}}$} &
\colhead{$N_{f}$} &
 \colhead{freq.} &
\colhead{Combination} \\
& &  \colhead{}  & \colhead{(d$^{-1}$)} &  & \colhead{(d$^{-1}$)} 
} 
{\startdata
		 29984014 	&	 HD 268687 	&	 6 	&	 2.765 	&	 13 	&	 $f_0 = 2.76526$\\[1.5ex]
		 31106686 	&	 HD 33579 	&	 13 	&	 0.177 	&	 1 	&	 $f_0 = 0.17713$\\[1.5ex]
		 31109182 	&	 HD 268946 	&	 8 	&	 0.168 	&	 6 	&	 $f_0 = 0.16846$\\[1.5ex]
		 179304909 	&	 SK -69 99 	&	 2 	&	 0.292 	&	 5 	&	 $f_0 = 0.29210$\\[1.5ex]
		 276863889 	&	 HD 269787 	&	 2 	&	 0.909 	&	 9 	&	 $f_0 = 0.90868$\\[1.5ex]
		 276864037 	&	 HD 269781 	&	 13 	&	 0.072 	&	 3 	&	 \dots \\[1.5ex]
		 276869010 	&	 HD 269762 	&	 4 	&	 0.176 	&	 7 	&	 $f_{0} = 0.17614$\\[1.5ex]
		 277108449 	&	 HD 269840 	&	 3 	&	 0.717 	&	 13 	&	 $f_{0} = 0.71703$\\[1.5ex]
		 279956577 	&	 HD 269604 	&	 1 	&	 0.105 	&	 10 	&	 $f_{0} = 0.10522$ 	&	 \\[0.5ex]
								 	&	  	&	 	&	 	&	  	&	 $f_{1} = 0.09012$ 	&	\\ 
								 	&	  	&	 	&	 	&	  	&	 $f_{2} = 0.19419$ 	&$f_{0}+f_{1}$ \\[1.5ex]
		 279957325 	&	 CD-69 310 	&	 13 	&	 0.101 	&	 10 	&	 $f_{0} = 0.10070$\\[1.5ex]
		 391813303 	&	 HD 269651 	&	 2 	&	 0.098 	&	 4 	&	 $f_{0} = 0.09828$\\[1.5ex]
		 391815407 	&	 HD 269661 	&	 21 	&	 4.784 	&	 15 	&	 \dots \\[1.5ex]
		 404850274 	&	 HD 269953 	&	 6 	&	 2.671 	&	 21 	&	 $f_{0} = 2.67059$\\[1.5ex]
		 425083216 	&	 HD 269723 	&	 2 	&	 2.374 	&	 6 	&	 $f_{0} = 2.37353$\\[1.5ex]
         \hline
         181043309 	&	 HV 829 	&	 30 	&	 29.893 	&	 5 	&	 \dots \\[1.5ex]
		 181446366 	&	 [VA82] II-2 	&	 1 	&	 0.899 	&	 6 	&	 $f_{0} = 0.90173$ 	&	 \\[0.5ex]
								 	&	  	&	 	&	 	&	  	&	 $f_{1} = 1.80015$ 	&	 $2\,f_{0}$\\[1.5ex]
		 267547804 	&	 BZ Tuc 	&	 4 	&	 0.183 	&	 1 	&	 $f_0 = 0.18212$\\[1.5ex]
		 \enddata }
		  \tablecomments{The TIC IDs of the stars are listed, followed by their common name. The first 14 stars are LMC targets, while the last three are from the SMC. The number of prewhitened frequencies identified by \citetalias{Dorn-Wallenstein2022} 
        is denoted as $N_f^{\rm DW}$, while $f_0^{\rm DW}$ is the corresponding highest amplitude frequency that they recover and do not attribute to a contaminating star. $N_f$ is our number of prewhitened frequencies. The sixth column lists the closest frequency from our list of extracted frequencies that match with the corresponding $f_0^{\rm DW}$ within the Rayleigh limit, as well as related frequencies used in our \tl\ analysis. Empty entries means that no frequencies matched with $f_0^{\rm DW}$ within the Rayleigh limit. For the stars with multiple frequencies listed, the last column denotes how they are related.}
\end{deluxetable*}
}

In 3 out of 17 cases, we were unable to identify a significant periodogram peak within the Rayleigh limit ($\sigma_R = 1/T$) of the tabulated $f_0$ values. We attribute this to be due to differences in our adopted prewhitening and frequency filtering procedures. For HD~269781, the $f_0$ signal was excluded from our initial extracted frequency list when filtering for close frequencies due to a higher amplitude signal being present at 0.0648\,d$^{-1}$. By lowering our S/N stopping criterion, the $f_0$ signal for HD~269661 could be extracted (4.78414\,d$^{-1}$, $S/N = 2.963$), while a signal at 29.85982\,d$^{-1}$ ($S/N = 2.40$) within $1.65\, \sigma_R$ of $f_0$ could be found for HV~829.

We identify signals that are arithmetically related to $f_0$ (e.g., integer multiples, sums or differences) as either belonging to the same harmonic sequence or a set of combination frequencies. These signals are likely to be physically associated with the same source, as it is unlikely for unrelated signals to fall within frequency bins that are related in these ways by chance given the high frequency resolution of TESS (finer than 0.003\,d$^{-1}$ for stars in the TESS continuous viewing zones). The crossmatched $f_0$ values and associated combination frequencies used for \tl\ are listed in Table~\ref{Tab:f0}. 

Of the fourteen stars for which we can confirm a significant peak at $f_0$, we obtain at least one fit meeting our quality criteria for six. Four of these have consistent source locations returned for multiple sectors, though the signal reported for BZ Tuc appears not to be astrophysical upon further scrutiny. We summarize the results for each of the six stars below.

\subsection{HD~269953}

HD~269953 was analyzed in detail in Section~\ref{sec:case_example}. The $f_0$ signal and its subharmonic are offset from the YSG target by 41.5\arcsec\ and coincide with the O-type star BI~265. Localization of these signals met our reliability criteria in 22 out of 23 available sectors of TESS data.

\subsection{HD~268687}

\begin{figure*}
\center
\includegraphics[width=0.90\linewidth]{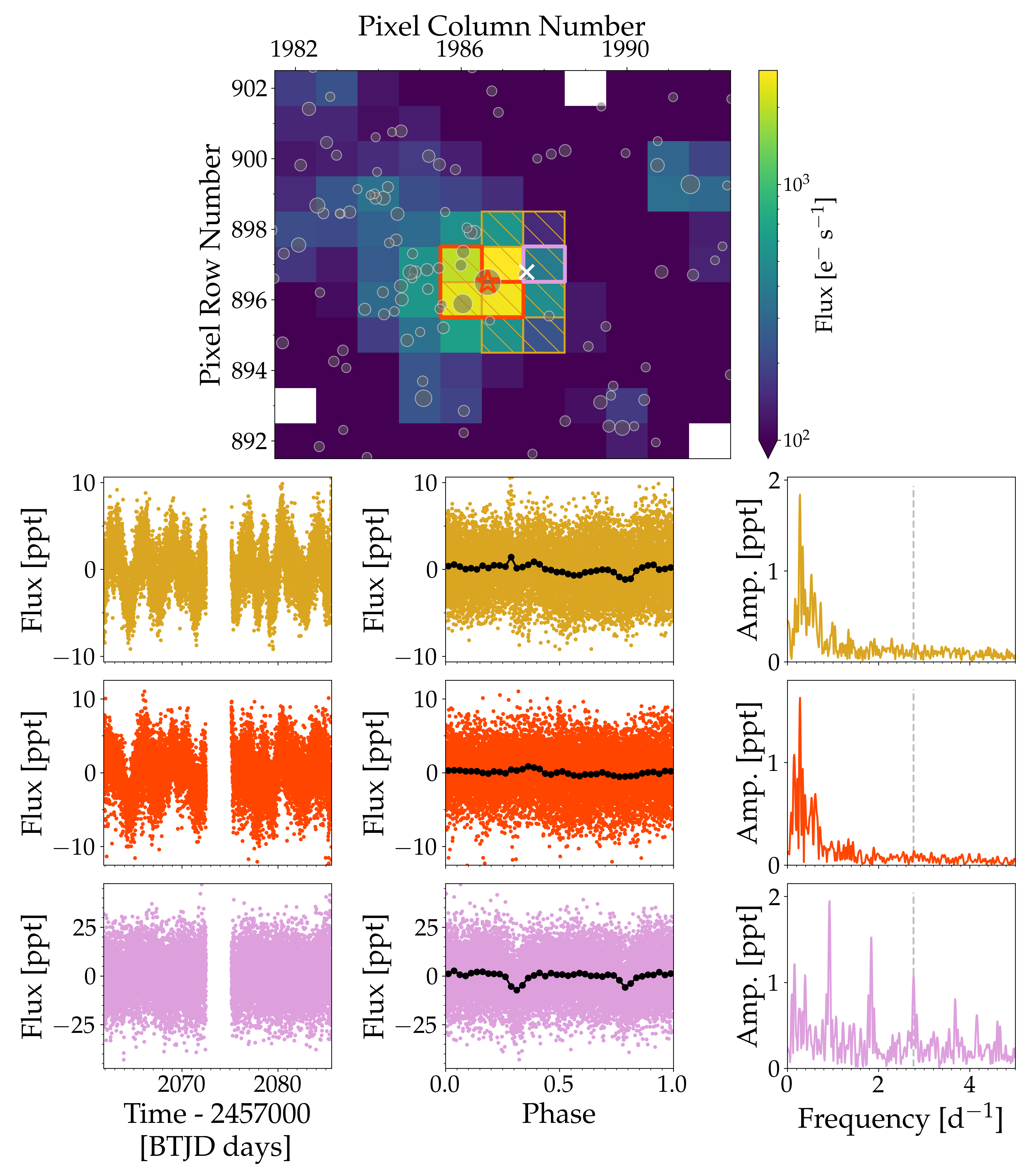}
\caption{Median TPF flux of TESS sector 28 for the YSG star HD~268687 (TIC~29984014, red star) and position of a nearby contaminating star indicated by the white cross. Bottom panels show the extracted light curve (left), the phase folded light curves using $\frac{1}{6}f_0 = 0.460877$\,d$^{-1}$ (center), and the periodograms (right) for the three different pixel masks indicated in the top panel. The $f_0$ value of the the highest amplitude independent signal identified by \citetalias{Dorn-Wallenstein2022} based on the SPOC aperture (yellow) is shown by vertical dashed line in the periodograms.}\label{fig:HD268687}
\end{figure*}

\begin{figure}
\center
\includegraphics[width=\linewidth]{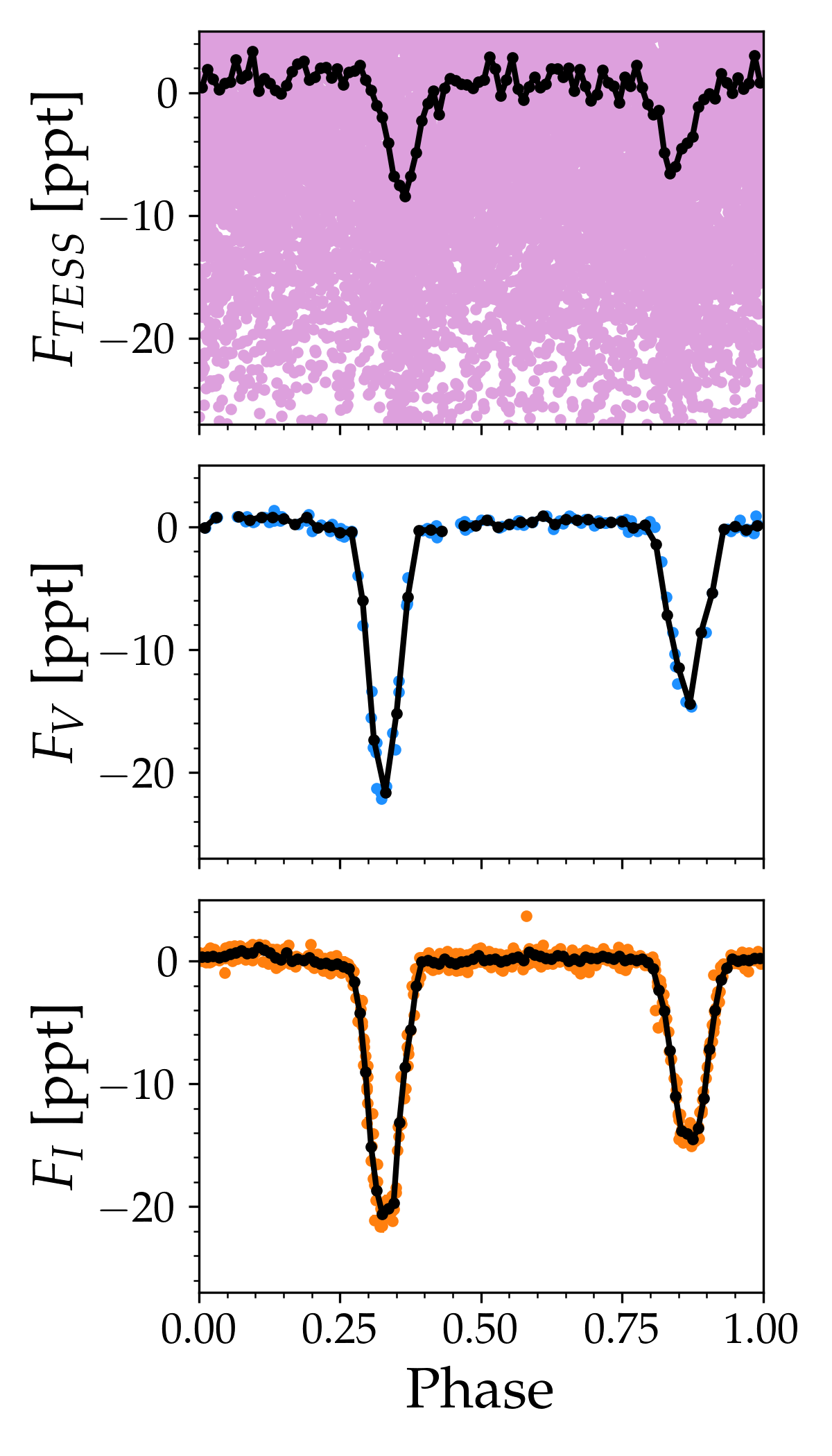}
\caption{Phase binned light curves of OGLE~LMC-ECL-1544 using the 2-min cadence TESS data (top panel; Same as bottom center panel of Fig.~\ref{fig:HD268687}), and the OGLE V (middle) and I (bottom) band photometry 
\citep{Udalski1997AcA....47..319U,Szymanski2005AcA....55...43S}. The light curves were phases folded assuming $f_{\rm orb} = \frac{1}{6}f_0 = 0.460877$\,d$^{-1}$. The black curves show the average flux in 0.01 (top), 0.02 (middle), and 0.01 (bottom) phase bins.}\label{fig:HD268687_OGLE}
\end{figure}

HD~268687 ($V=10.73$\,mag) returned good fits for 20 out of 25 sectors, all consistent with an average location of (72.7433, -69.4356) degrees, which is 20.5\arcsec\ from the target, see Fig.~\ref{fig:HD268687}. This position is most consistent with a known eclipsing binary, OGLE~LMC-ECL-1544 ($V=16.043$\,mag), with an orbital period of 2.169763 days \citep{2011AcA....61..103G}, of which $f_0$ is the 6th harmonic. The second column of panels in Fig.~\ref{fig:HD268687} shows the three extracted light curves from SPOC, the target, and the identified source of the $f_0$ signal, phase folded by $\frac{1}{6}f_0$. The binned averages shown in black reveal the eclipses of the OGLE~LMC-ECL-1544 binary, which are otherwise difficult to see due to the noisy TESS data. For comparison, we show in Fig.~\ref{fig:HD268687_OGLE} the phase folded TESS light curve (top panel) and the corresponding phase-folded OGLE $V$ and $I$-band light curves (two bottom panels) of the eclipsing binary, where the eclipses show up much more clearly in the OGLE data. As TESS is designed for observing bright targets, this difference is not unexpected given the faintness of the binary. Nevertheless, we note in spite of the contaminating star being more than five magnitudes fainter than the target YSG, the eclipsing binary variability is still influencing the SPOC light curve.

\subsection{HD~33579}

HD~33579 has 12 sectors of TESS data available and \tl\ reported a marginally significant result for only Sector 29. The data from that sector shows considerable systematic noise, which can confuse localizations, especially for such low-frequency signals that only complete a few cycles per sector. Combined with a lack of corroboration from other sectors, we interpret this single ``good'' fit as spurious.

\subsection{HD~268946}

HD~268946 returns three localizations that pass our initial quality criteria out of twelve; however, we reject them as spurious detections since they are found at inconsistent sky locations. Inspecting the light curves suggests that the fits are coupling to TESS systematics with similar timescales as the alleged signal.

\subsection{HD~269787}

\begin{figure*}
\center
\includegraphics[width=0.90\linewidth]{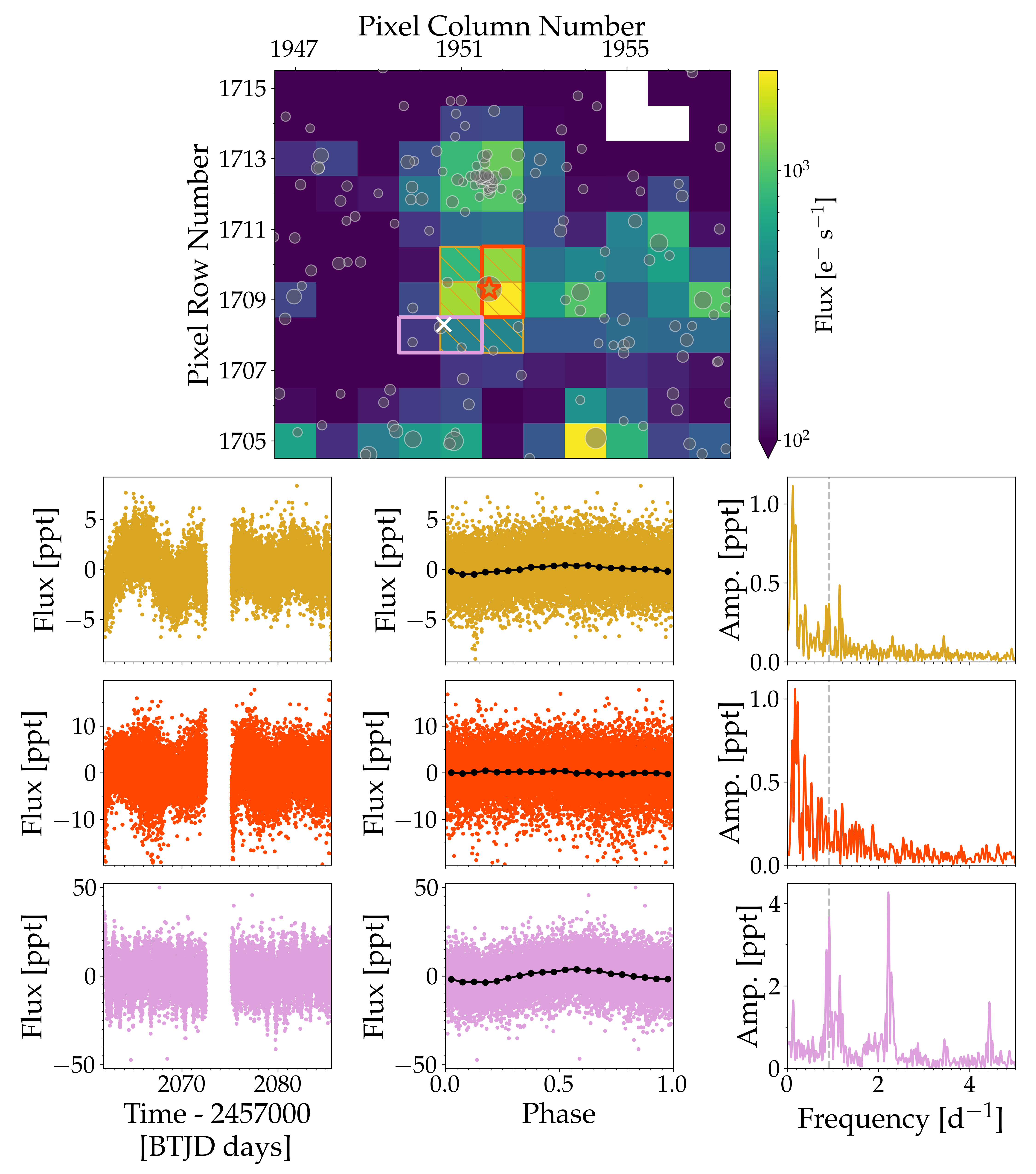}
\caption{Median TPF flux of TESS sector 28 for the YSG star HD~269787 (TIC~276863889, red star) and position of a nearby contaminating star indicated by the white cross. Bottom panels show the extracted light curve (left), the phase folded light curves using $f_0 = 0.90868$\,d$^{-1}$ (center), and the periodograms (right) for the three different pixel masks indicated in the top panel. The $f_0$ value of the the highest amplitude independent signal identified by \citetalias{Dorn-Wallenstein2022} based on the SPOC light curves (yellow) is shown by the vertical dashed line in the periodograms.}\label{fig:HD269787}
\end{figure*}

HD~269787 ($V=10.73$\,mag) data yield six good localizations out of ten available sectors of data, all at consistent sky locations. The source of the $f_0$ signal is centered on (83.6215,-66.9711) degrees on average, which is 31.5\arcsec\ from the position of the YSG target as shown in Fig.~\ref{fig:HD269787}.  The reported signal can be uniquely ascribed to the object AL~327 (TIC~276863886, $V=14.80$\,mag), an emission-line star \citep{1964IrAJ....6..241A,2013A&A...555A.141H} not previously studied as a photometric variable. We note that the contaminating star is three magnitudes fainter than the YSG.

\subsection{BZ Tuc}

\begin{figure}
\center
\includegraphics[width=\linewidth]{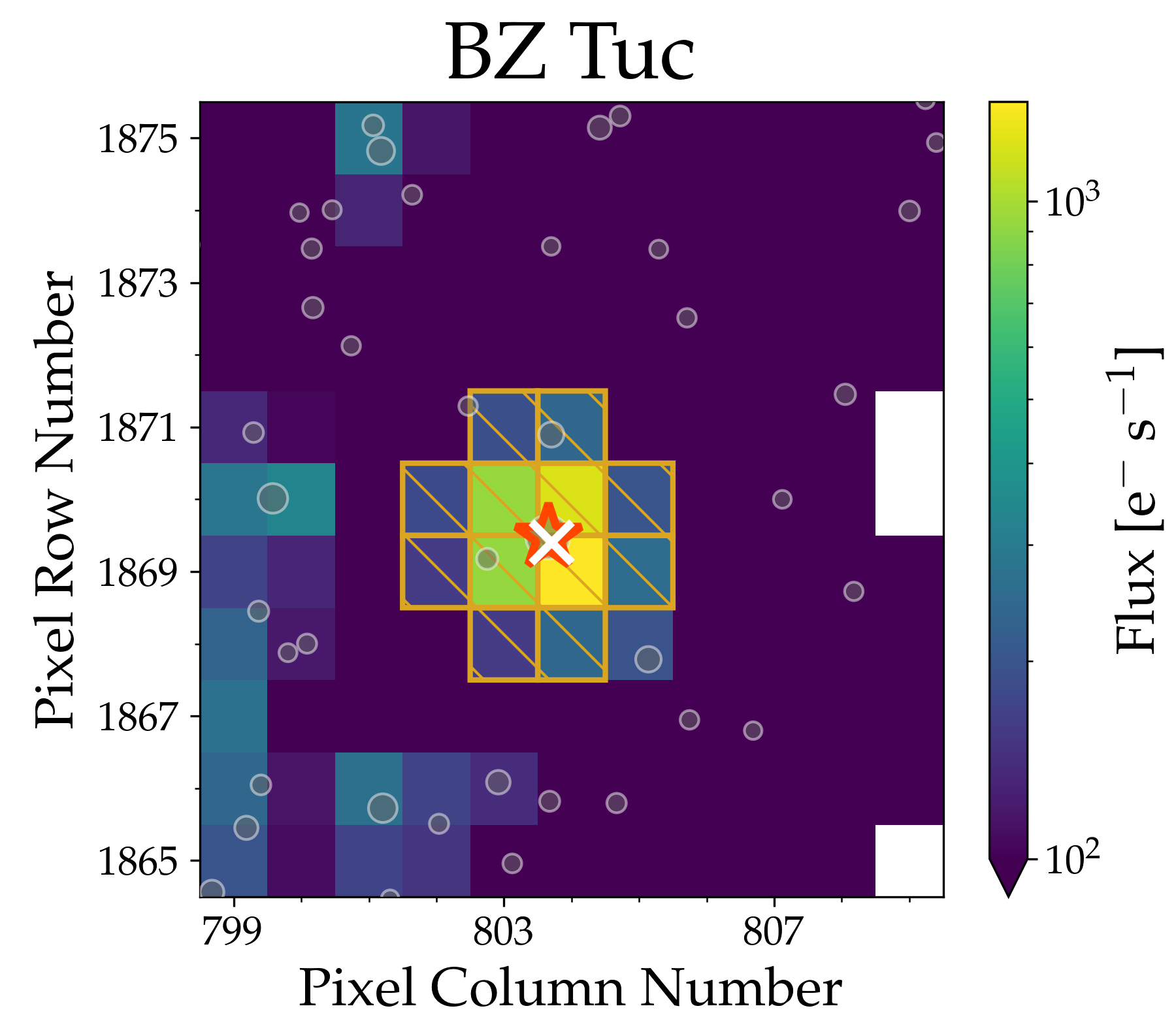}
\caption{
TESS localization result for BZ~Tuc (TIC~267547804), showing that the $f_0$ signal identified by \citetalias{Dorn-Wallenstein2022} coincides with the position of the YSG on the TPF. Positions of nearby Gaia sources brighter than 17\,mag are also shown.
}\label{fig:BZ_Tuc}
\end{figure}

BZ Tuc has two sectors of data available, and \tl\ localized the main reported signal to the target in both, as shown in Fig.~\ref{fig:BZ_Tuc}. 
This is a fairly well studied Cepheid variable with one of the longest known periods of 127 days \citep[as HV 821, e.g.,][]{1965MNRAS.130..333G,1977ApJS...34....1E}. In general, it is difficult to reliably measure intrinsic periods of variation close to or exceeding the TESS orbital period of 13.7 days, as this is a typical timescale of TESS systematics, and TESS light curves show significant systematic differences in flux zero points between sectors. This limitation of the TESS data has been borne out in efforts to measure stellar rotation periods \citep{2020ApJS..250...20C,2022ApJ...930....7A,2022ApJ...936..138H}.

\begin{figure*}
\center
\includegraphics[width=\linewidth]{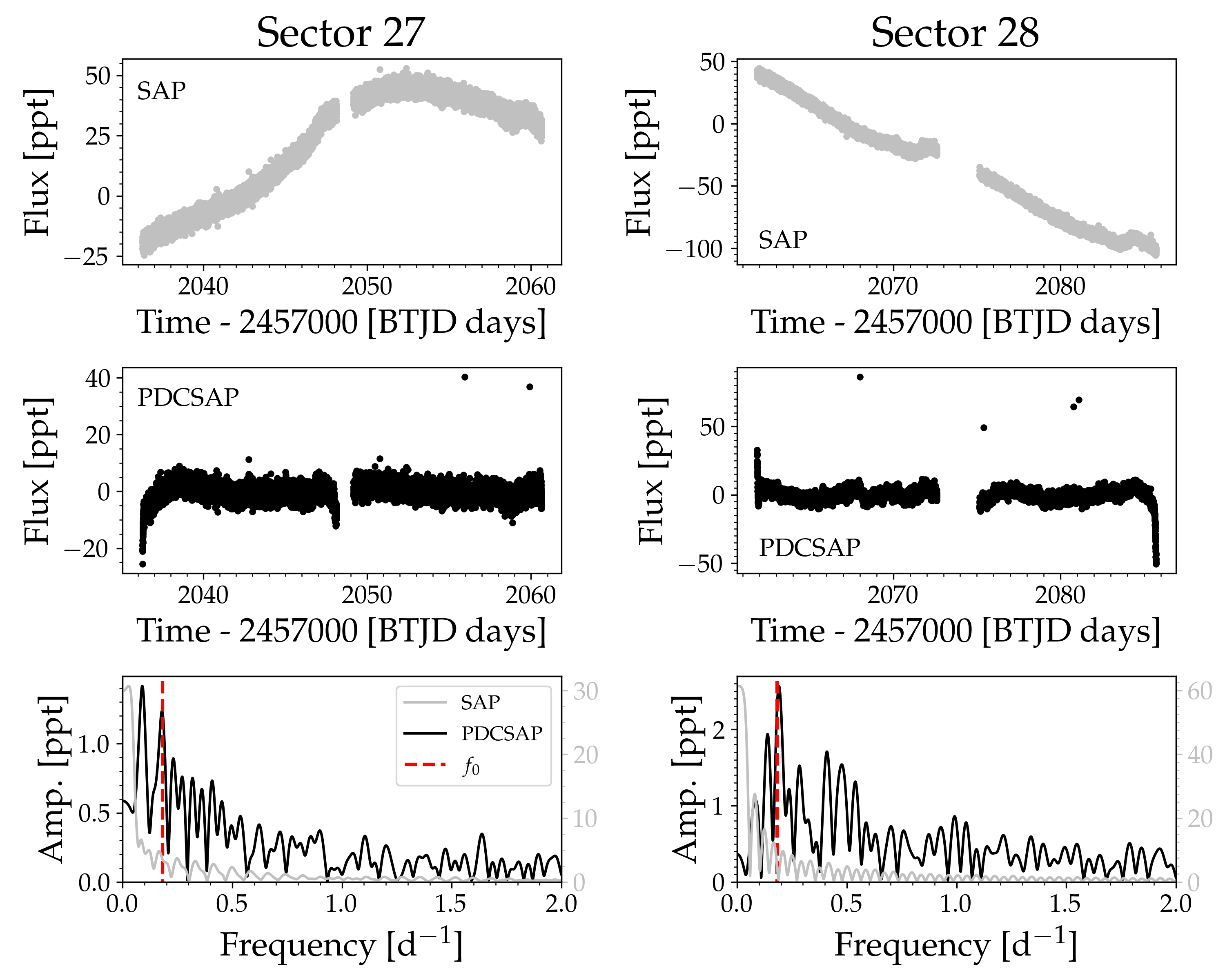}
\caption{SAP (top panels, grey) and PDCSAP (middle panels, black) light curves of BZ~Tuc for the two available 2-min cadence sectors. The bottom panels show the periodogram of PDCSAP and SAP light curves with the $f_0$ signal indicated by the dashed red line. Due to the different amplitude scales of the PDCSAP and SAP periodograms, the SAP amplitudes are shown in grey on the right-hand sides of both bottom subplots in units of ppt.}
\label{fig:BZ_Tuc_sap_pdcsap}
\end{figure*}

Figure~\ref{fig:BZ_Tuc_sap_pdcsap} shows both types of light curves produced for BZ Tuc by the TESS SPOC pipeline for sector 27 (left) and 28 (right): the simple aperture photometry (SAP) light curves that are directly extracted from the target aperture (top), and the PDCSAP light curves that have been further processed in an attempt to mitigate observational systematics (middle). These light curves were normalized separately for each sector by dividing by the median flux in the sector, centering around zero, and then changing to units of ppt. The long-timescale Cepheid variability is visible in the SAP light curves, but conspicuously absent in the PDCSAP light curves, which are instead dominated by systematic errors. The additional light curve detrending of the PDC algorithm in the SPOC pipeline fits and removes combinations of cotrending basis vectors (CBVs)  from the SAP light curves. CBVs represent systematic trends that principal component analysis reveals to be most common among TESS light curves \citep{2012PASP..124..963K}. The PDCSAP light curves are what \citetalias{Dorn-Wallenstein2019}--\citetalias{Dorn-Wallenstein2022} worked with, and what we also analysed to reproduce their work. Observational systematics characterized by the CBVs often vary on similar timescales as the observing run duration or spacecraft orbit that can mimic long-timescale astrophysical signals. Fitting CBVs to such targets tends to over-fit the data, removing the astrophysical signals of interest while introducing or amplifying other systematics from the residuals of the CBV fitting. For these reasons, using the CBV-corrected light curves for long-period variables like BZ Tuc is not advised \citep{2012PASP..124..963K}. For the BZ Tuc light curves, the PDC pipeline opted to use the Multi-Scale correction method \citep{2014PASP..126..100S}, even though the Single-Scale option \citep{2012PASP..124.1000S} tends to better preserve long-timescale signals.

The bottom panels of Figure~\ref{fig:BZ_Tuc_sap_pdcsap} show the periodograms of the two PDCSAP (black) and SAP (grey) light curves separately for each sector, with the reported $f_0$ signal indicated (red dashed line). The $f_0$ only appears as a compelling peak in the PDCSAP light curves, which have been overcorrected by the TESS pipeline. But \tl\ extracts the light curves from each pixel, and it still localized this signal to BZ Tuc. As an optional step, \tl\ does consider applying its own PCA corrections to the extracted light curves, and in this case it recognized that the CBVs it computed contain significant power at the frequency of interest so it opted not to risk applying these corrections.  Still, the long-timescale variations observed in these pixels does include some power from the Cepheid pulsations (and potentially also systematics) at the spurious $f_0$ frequency---the gray periodograms of the SAP light curves show slightly elevated amplitudes at this frequency compared to the average noise level observed at higher frequencies. This power excess at $f_0$ from the longer-timescale Cepheid variability caused the signal to be localized to BZ Tuc despite it not being an unique independent pulsation signal of this star. This highlights one of the failure modes of \tl, due to the violation of a base assumption: that the input frequencies represent significant and unique signals from one source in the field.

\section{Conclusions}\label{sec:conclusions}

Of the first five YSGs that \citetalias{Dorn-Wallenstein2020} 
used to argue the existence of a new FYPS class of pulsating variable star, \citetalias{Dorn-Wallenstein2022} 
attribute the variability of HD~269110 and HD~269902 to contaminants based on OGLE data (\tl\ confirms). We have shown that the main signals from HD~269953 and HD~268687 also originate from other sources. \citetalias{Dorn-Wallenstein2022} found the 1.135\,day$^{-1}$ signal that \citetalias{Dorn-Wallenstein2020} 
identified as the strongest for the remaining candidate, HD~269840, to be caused by contamination, which we confirm with \tl. The remaining variations in the light curve that \citetalias{Dorn-Wallenstein2022} 
associate with HD~269840 are too weak for \tl\ to locate, but there is no reason to assume that they belong to the YSG in such a crowded part of the sky. 

Of the 14 YSG stars with a confirmed significant peak at the $f_0$ signals reported by \citetalias{Dorn-Wallenstein2022}, we find consistent source locations for four of them, including the previously mentioned HD~269953 and HD~268687 stars. For three of these, the $f_0$ signals are associated with nearby contaminating star, while the signal for known Cepheid variable BZ~Tuc is localized to the target. In the latter case, we find that the $f_0$ signal is likely a spurious signal introduced by the PDC algorithm of the SPOC pipeline to the PDCSAP light curve when removing the long-period Cepheid variability from the SAP light curve. Finally, while \citetalias{Dorn-Wallenstein2022} make a statistical argument that all claimed FYPS are unlikely to be caused by contamination, we argue that FYPS should not be considered a class of pulsating star until rapid pulsational variability can be uniquely associated with a YSG.

This work demonstrates how much of a challenge source confusion can be for analyses of TESS data. Even if contamination of the aperture is predicted to be low (estimated by the {\tt CROWDSAP} header value), significant signals can still be caused by source blending \citep{Higgins2022}. As seen in this work, signals of nearby variable stars can contaminate light curves even when they are more than five magnitudes fainter than the target stars. Interrogating TESS data at the pixel level for contamination issues is essential if misattribution of signals is to be avoided; the {\tt lightkurve} Python package can help to make the pixel data more accessible \citep{lightkurve}. Statistical arguments for the likelihood of contamination of TESS light curves may not be reliable.  For localizing (multi-)periodic sinusoidal variability to sub-pixel precision, the Python tool \tl\ can help in many cases \citep[for guidance, see][]{Higgins2022}.

\clearpage
\begin{acknowledgments}
\small
The authors are grateful to Jim Fuller, Lars Bildsten, Tim Bedding, Conny Aerts, and Trevor Dorn-Wallenstein for their valuable discussions and comments at different stages of the manuscript. We thank the referee for their comments which improved the manuscript. 
This research was supported in part by the National Science Foundation under Grant No. NSF PHY-1748958 as well as through the TESS Guest Investigator program Cycle 4 under Grant No.\ 80NSSC22K0743 from NASA and by the Professor Harry Messel Research Fellowship in Physics Endowment, at the University of Sydney.
This research benefited from interactions with Lars Bildsten and Jim Fuller that were funded by the Gordon and Betty Moore Foundation through Grant GBMF5076. 
This research made use of Lightkurve, a Python package for Kepler and TESS data analysis \citep{lightkurve}. This paper includes data collected with the TESS mission, obtained from the MAST data archive at the Space Telescope Science Institute (STScI). Funding for the TESS mission is provided by the NASA Explorer Program. STScI is operated by the Association of Universities for Research in Astronomy, Inc., under NASA contract NAS 5–26555. This research has made use of the VizieR catalogue access tool, CDS, Strasbourg, France. This work has made use of data from the European Space Agency (ESA) mission
{\it Gaia} (\url{https://www.cosmos.esa.int/gaia}), processed by the {\it Gaia}
Data Processing and Analysis Consortium (DPAC,
\url{https://www.cosmos.esa.int/web/gaia/dpac/consortium}). Funding for the DPAC
has been provided by national institutions, in particular the institutions
participating in the {\it Gaia} Multilateral Agreement.
\\
\end{acknowledgments}

%

\vspace{5mm}
\facilities{TESS \citep{TESS}, Gaia \citep{GaiaCollaboration2016A&A...595A...1G,GaiaCollaboration2022arXiv220800211G,Babusiaux2022arXiv220605989B}}


\software{\texttt{astropy} \citep{2013A&A...558A..33A,2018AJ....156..123A},  
          \texttt{TESS\_localize} \citep{Higgins2022}, 
          \texttt{lightkurve} \citep{lightkurve},
          \texttt{astroquery} \citep{Ginsburg2019},
          \texttt{tesscut} \citep{Brasseur2019},
          \texttt{numpy} \citep{harris2020array}, 
          \texttt{matplotlib} \citep{Hunter2007},
          \texttt{pandas} \citep{reback2020pandas,mckinney-proc-scipy-2010},
          \texttt{MESA} \citep{Paxton2011,Paxton2013,Paxton2015,Paxton2018,Paxton2019,Jermyn2022arXiv220803651J}.
          }



\appendix

\section{\texttt{MESA} setup}\label{Sec:mesa}

For the construction of Fig.~\ref{fig:acou_cutoff} we computed non-rotating stellar evolution tracks for five different initial stellar masses (15\,M$_\odot$, 20\,M$_\odot$, 25\,M$_\odot$, 30\,M$_\odot$, 35\,M$_\odot$) using the stellar structure and evolution code \texttt{MESA} version 22.11.1
\citep{Paxton2011,Paxton2013,Paxton2015,Paxton2018,Paxton2019,Jermyn2022arXiv220803651J}. The models were evolved from the Hayashi track to the end of core He burning, assuming an initial stellar mass fraction for hydrogen, helium, and metals of $X=0.7391$, $Y=0.2562$, and $Z=0.0047$ for the LMC following \cite{Kohler2015}, who based these values on the observed abundances of young massive stars in the LMC \citep{Brott2011}. The \cite{Asplund2009} metal mixture and corresponding OP opacity tables were adopted. We used the Ledoux criterion with time-dependent convection, and adopted a small amount of convective boundary mixing following the exponential diffusive overshoot formalism \citep{Freytag1996,Herwig2000}. Finally, we included mass loss from the \cite{Vink2001} hot wind scheme assuming a scaling factor of 0.3 \citep{Bjorklund2021}. The full \texttt{MESA} inlist is available at DOI:\dataset[10.5281/zenodo.7773563]{https://doi.org/10.5281/zenodo.7773563}.

\section{Light curve extraction and correction for the HD~269953 example}\label{Sec:lc_extract}

The light curves and associated periodograms shown in Fig.~\ref{fig:Cont_example1} were obtained in two different ways. For the SPOC target pixel mask indicated by the hatched yellow region in the top panel of the figure, we downloaded the SPOC 2-min cadence PDCSAP flux light curve from MAST using the \texttt{lightkurve} Python package. The downloaded light curve was subsequenctly normalized by dividing by the median flux of the sector, and the flux units changed to parts-per-thousand (ppt). Finally, the light curve is centered around zero and the Lomb-Scargle periodogram calculated. These median normalized SPOC PDCSAP flux light curves correspond to the ones analyzed by \cite{Dorn-Wallenstein2019,Dorn-Wallenstein2020,Dorn-Wallenstein2022}.

For the remaining four pixel masks, the light curves were extracted and corrected manually once again relying on \texttt{lightkurve} and its \texttt{RegressionCorrector} class. The  four light curves were extracted using simple aperture photometry for the selected target pixel mask of HD~269953 (red) and three pixel masks for the three contaminating stars shown in grey, pink, and blue in Fig.~\ref{fig:Cont_example1}. The background flux was extracted by creating a background pixel mask, where all pixels fainter than 0.001 times the standard deviation above the overall median flux in the TPF were included. To subsequently reduce the dimensionality of the design matrix of the background flux used for the linear regression, a principal component analysis was performed where the dominant flux variability of the background pixels were captured using up to five principal components. 

Using the constructed design matrix, the extracted target and contaminator light curves are corrected by creating a model of the background flux using linear regression and subtracting the model from the uncorrected light curves. Finally, the corrected light curves are normalized using a low order polynomial (typically 1st order but up to 3rd depending on the sector) and centered around zero after changing the flux units to ppt. The final light curves and associated periodograms for Sector 34 are shown in the six bottom panels of Fig.~\ref{fig:Cont_example1}. The same methodology was adopted to extract and correct the light curves shown in Figs.~\ref{fig:HD268687} and \ref{fig:HD269787}.






\end{document}